
\global\overfullrule 0pt
\magnification = \magstep1
\newcount\fignumber
\newcount\notenumber
\newcount\bibno
\newcount\eqnumber

\def\sec#1{\centerline{{\bf#1}}}

\def\dd{\partial}

\def\s{\ifmmode \widetilde \else \~\fi}
\def\lta{\mathrel{\rlap{\lower 3pt\hbox{$\mathchar"218$}}
    \raise 2.0pt\hbox{$<$}}}
\def\gta{\mathrel{\rlap{\lower 3pt\hbox{$\mathchar"218$}}
    \raise 2.0pt\hbox{$>$}}}

\def\num{\vskip 6pt\global\advance\bibno by 1\item{\number\bibno .\ }}
\def\note{\global\advance\notenumber by 1\footnote{$^{\the\notenumber}$}}
\def\new{(\chaphead\the\eqnumber\global\advance\eqnumber by 1 }
\def\last{\advance\eqnumber by -1 (\chaphead\the\eqnumber
    \advance\eqnumber by 1 }
\def\eq#1{\advance\eqnumber by -#1 equation (\chaphead\the\eqnumber
    \advance\eqnumber by #1 }
\def\eqnam#1{\xdef#1{\the\eqnumber}}
\def\nfig{\chaphead\the\fignumber\ \global\advance\fignumber by 1 }
\def\nfiga#1{\chaphead\the\fignumber{#1}\global
    \advance\fignumber by 1 }
\def\rfig#1{\advance\fignumber by -#1 \chaphead\the\fignumber
    \advance\fignumber by #1 }
\def\chaphead{}

\hyphenpenalty=1000
 at 12 truept

\def\pmb#1{\setbox0=\hbox{#1}%
\kern-.025em\copy0\kern-\wd0
\kern.05em\copy0\kern-\wd0
\kern-.025em\raise.0433em\box0 }
\def\del{\hbox{\pmb{$\nabla$}}}
\def\btimes{\hbox{\pmb{$\times$}}}
\def\bcdot{\,\hbox{\pmb{$\cdot$}}\,}

\def\thismonth{\ifcase\month\or January\or February\or March\or April\or May\or
June\or July\or August\or September\or October\or November\or December\fi}
\def\date{\number\day\ \thismonth\ \number\year}
%\headline={ \hfil -- {\twelverm\folio} --\hfil}% DRAFT: DO NOT CIRCULATE}

%\headline={ {\it Version:} \date\hfil -- {\twelverm\folio} --\hfil 
%DRAFT: DO NOT CIRCULATE}
%\def\pn{\par\noindent}

%\input sokhead
\def\bv { {\bf v} }
\def\bw { {\bf w} }
\def\eps { {\epsilon} }
\hfuzz= 10 pt
\parskip=6pt

\baselineskip = 24pt
%\hoffset=0.25 truein

\font\nice=cmcsc10 scaled \magstep1
\null
\vskip 0.25truein
\centerline{\bf ON NONSHEARING MAGNETIC CONFIGURATIONS}
\centerline {\bf IN DIFFERENTIALLY ROTATING DISKS.}
\centerline{by}
\centerline{\nice Steven A. Balbus}
\baselineskip 12pt
\smallskip
\centerline {Virginia Institute for Theoretical Astronomy,}
\centerline {Department of Astronomy,}
\centerline {University of Virginia,}
\centerline {Charlottesville, VA 22903-0818}
\medskip
\centerline{and}
\medskip
\centerline{\nice Massimo Ricotti}
\smallskip 
\centerline {Dipartimento di Astronomia}
\centerline {Universit\'a di Firenze,}
\centerline {Largo E. Fermi 5}
\centerline {I-50125, Firenze, Italy}
\smallskip 
\centerline {Center for Astrophysics and Space Astronomy}
\centerline {University of Colorado,}
\centerline {Campus Box 389}
\centerline {Boulder, CO 80309-0389}
\vskip 48pt
\par\noindent{\sl Received:}
\vskip 24pt
\par\noindent Version: \date
\baselineskip 12pt
\vskip 2 truein
\vfill\eject

%\input sokmacros2
%\input sokhead
%\def\P{  {\cal P}  }
%\def\bv { {\bf v} }
%\def\eps { {\epsilon} }
%\pageno=3
%\raggedright\raggedbottom
%\baselineskip =24pt
%\parskip =5 pt
%\hfuzz= 10 pt
%\def\tilde{\widetilde}

\pageno=2
\sec{\bf ABSTRACT}

A new class of disk MHD equilibrium solutions is described, which is
valid within the standard local (``shearing sheet'') approximation
scheme.  These solutions have the following remarkable property:
velocity streamlines and magnetic lines of force rotate rigidly, even
in the presence of differential rotation.  This situation comes about
because the Lorentz forces acting upon modified epicycles compel fluid
elements to follow magnetic lines of force.  Field line (and
streamline) configurations may be elliptical or hyperbolic, prograde or
retrograde.  These structures have previously known hydrodynamical
analogs: the ``planet'' solutions described by Goodman, Narayan, \&
Goldreich.

The primary focus of this investigation is configurations in the disk
plane.  A related family of solutions lying in a vertical plane is
briefly discussed; other families of solutions may exist.  Whether
these MHD structures are stable is not yet known, but could readily be
determined by three-dimensional simulations.  If stable or
quasi-stable, these simple structures may find important applications in
both accretion and galactic disks.  \vfill\eject

\sec{\bf 1. INTRODUCTION}
\smallskip

The essential role of magnetohydrodynamics (MHD) in understanding the
behavior of accretion disks is now widely recognized (e.g. Papaloizou
\& Lin 1995, Balbus \& Hawley 1998).  Usually, however, detailed
solutions to the dynamical equations are quite complicated:  one must
contend with turbulence in weak field systems, and exact equilibrium
solutions, when they can be found at all, are highly technical and
rather special.  Generally, differential rotation will complicate
matters by producing a time-dependent toroidal magnetic field when a
radial field is present.

Given the above, it might appear that any true equilibrium MHD
configuration would require at the very least a vanishing radial field
component.  This is not so.  In this {\it Letter}, we present exact,
two-dimensional solutions to the MHD equations, using only the
approximation of the standard shearing sheet model.   (In essence,
curvature effects are neglected).  The solutions are noteworthy for the
fact that the field lines, which do indeed have radial components,
rotate rigidly, even if the disk is not rotating uniformly.
Furthermore, the field configurations have very simple geometries:
hyperbolae and ellipses.  The dynamical and induction equations are
mutually consistent because magnetic forces cause departures from the
standard epicyclic paths, departures which in these solutions lead
fluid elements directly along the field lines.  Gas pressure gradients
are unimportant.

Of dynamical interest on their own, these solutions may also be
astrophysically relevant.  Since only the shearing sheet approximation
is used, both galactic as well as Keplerian disks are possible venues.
The crucial question is whether the solutions are stable--or more
generously, whether they are {\it stable enough.\/}  We do not attempt
to answer the stability question in this paper, which can be best
addressed by three-dimensional numerical simulations.  It should be
noted, however, that the configurations are not obviously unstable to
the weak field instability discussed by Balbus \& Hawley (1991), since
the field strengths involved fall at the edge of the stability domain.
If they are long lived, the elliptical solutions (which can be both
prograde and retrograde), would represent coherent disk structures.
One might then speculate that in galactic disks these magnetically
penned regions in galactic disks become natural sites for molecular
cloud complex formation.  But whether nature makes use of these
solutions or not is at present unknown.

Although very different in their detailed physics, these MHD solutions
are kin to the shearing sheet planet solutions discovered by
numerically by Hawley (1987), and elucidated analytically by Goodman,
Narayan, \& Goldreich (1987, hereafter GNG).  (The GNG solutions
allowed for the presence of significant pressure gradients in the
underlying equilibrium disk; the appropriate limit for comparison with
this work is restricted to Keplerian rotation profiles.) To understand
the similarity, note that the gravitational tidal force varies linearly
with radius (to lowest retained order).  This drives a velocity
response also linear in spatial coordinates.  The associated
streamlines are in both cases the familiar conic sections.  What is
surprising, however, is that the presence of Lorentz forces can be so
easily accommodated by this remarkably simple scaling.

In \S\ 2, we present a detailed derivation of the MHD structures,
and in \S\ 3 a brief discussion of the astrophysical implications
of these solutions is presented.

\bigskip
\sec {\bf 2. COHERENT MHD STRUCTURES}
\smallskip
\centerline{2.1 \it{ Basic Equations}}
\smallskip

Consider a differentially rotating disk, with
a cylindrical coordinate
system $(R, \phi, z)$ centered on the origin. 
The disk's angular velocity is given by $\Omega(R)$.
The fundamental MHD equations are mass conservation
$$
{d\ln\rho\over dt} + \del\bcdot{\bf w} = 0,
\eqno (2.1)
$$
the dynamical equation,
$$
\rho {d\bw\over dt} = -\del\left( P+ {B^2\over 8\pi}\right) -\rho
\del\Phi +({ {\bf B}\over4\pi} \bcdot \del){\bf B},
\eqno (2.2)
$$
and the induction equation,
$$
{\dd {\bf B}\over \dd t} = \del\btimes (\bw \btimes {\bf B})
%{d{\bf B}\over dt} = -{\bf B}\del\bcdot \bv +
%\del\btimes \left( \bv \btimes {\bf B} \right)
\eqno (2.3)
$$
Here, $d/dt$ is a Lagrangian time
derivative:
$$
{d\ \over dt} = {\dd\ \over \dd t} + {\bf w}\bcdot\del,
$$
$\bw$ is the velocity in the inertial frame,
$P$ represents the gas pressure, $\Phi$ is the central gravitational
potential, 
and the other symbols have their standard meanings. 
We shall not require an internal energy equation in our analysis.

We work in the local (or `shearing sheet') approximation.
This consists of choosing a fiducial radius $R_0$ rotating at
angular velocity $\Omega_0$, and erecting a local Cartesian system:
$$
x= R-R_0, \qquad y=R_0(\phi-\Omega_0 t), \qquad z=z.
\eqno (2.4)
$$
In the local approximation, $x\ll R_0$, and $v_x, v_y \ll R_0 \Omega_0$,
where the $v_i$ are velocities relative to a corotating origin.
The Alfv\'en velocity is defined by
$$
{\bf v_A} = { {\bf B}\over\sqrt{ 4 \pi \rho}}. 
$$ 
We shall assume initially assume that
the Alfv\`en velocity components $v_{Ax}$ and $v_{Ay}$ are $\ll R_0
\Omega_0$, a result which will prove to be self-consistent.
When present, $v_z$ and $v_{Az}$ will also be assumed to satisfy this
inequality.

Let the global scale angular velocity given by a power law of the form
$\Omega(R) = \Omega_0(R_0/R)^q$.  Keplerian flow corresponds to
$q=3/2$, a flat galactic rotation curve to $q=1$.  
Effecting our coordinate transformation (2.4) and letting $\bv$ now
represent the velocity relative to our new axes leads to the
{\it Hill equations\/} (e.g. GNG; Hawley, Gammie, \& Balbus 1995):
$$
{d\ln\rho\over dt} + \del\bcdot{\bf v} = 0,
\eqno (2.5a)
$$
$$
%\eqalignno{
%{d\bv\over dt} &= - {1\over\rho} \del \left( P + {B^2\over 8 \pi}\right)
%\cr
%&\quad + { {\bf B}\bcdot\del {\bf B} \over4\pi\rho } -2
%{\bf\Omega}\btimes\bv + 2q\Omega^2 x {\bf {\hat{e}_x} } - \Omega^2
%{\bf {\hat{e}_z} },& (2.5b)\cr
%}
{d\bv\over dt} = - {1\over\rho} \del \left( P + {B^2\over 8 \pi}\right)
 + { ({\bf B}\bcdot\del) {\bf B} \over4\pi\rho } -2
{\bf\Omega}\btimes\bv + 2q\Omega^2 x {\bf {\hat{e}_x} } - \Omega^2z
{\bf {\hat{e}_z}},
$$
$$
{\dd {\bf B}\over \dd t} = \del\btimes(\bv\btimes {\bf B} ),
\eqno (2.5c)
$$
where $v_\phi$ is now measured
relative to $R\Omega_0$, and all differential
operators may be taken to be Cartesian.  We have dropped the $0$
subscript from $\Omega_0$ in equation (2.5b).  

\bigskip
\centerline{2.2 \it {Solutions}}

As our first case, 
let $\rho$ and $P$ be functions of $z$ only, which is certainly 
consistent with the local approximation.  We seek solutions to
equation (2.5) in which divergence-free
fluid velocities and the magnetic field are
confined to the $xy$ plane, and ${\bf v_A}$ depends upon $x$ and $y$
only.
In component form, the
steady-state mass and induction equations may be written
$$
\dd_i v_i = 0,
\eqno (2.6a)
$$
$$
v_i\dd_i v_{Aj} = v_{Ai}\dd_i v_j,
\eqno (2.6b)
$$
with $i$ and $j$ taking on the values $x$ and $y$.  The summation
convention on repeated indices is used unless otherwise stated. 
The dynamical equations become
$$
v_i\dd_i v_x -2\Omega v_y = -\dd_x \left({v_A^2\over2}\right)
+ v_{Ai}\dd_i v_{Ax} +2q\Omega^2 x,
\eqno (2.6c)
$$
$$
v_i\dd_i v_y +2\Omega v_x = -\dd_y
\left({v_A^2\over2}\right) + v_{Ai}\dd_i v_{Ay},
\eqno (2.6d)
$$
$$
\dd_z\left( H+{\Omega^2z^2\over2}\right) =0.
\eqno (2.6e)
$$
where the enthalpy $H$ is defined by $dH=dP/\rho$.
The notation $\dd_i$ denotes the partial derivative with respect to the
$i$th Cartesian coordinate. 

The $z$ equation may be immediately integrated,
$$
H- H_c = {\Omega^2z^2\over2}
\eqno (2.7)
$$
where $H_c$ is the central midplane value of the enthalpy.  The 
vertical structure decouples completely from the planar dynamics of
this problem, and we need not pursue it further.
\bigskip

\centerline{2.2.1 \it{Hydrodynamic Limit.} }

To orient ourselves, let us recover some familiar solutions for
nonmagnetized disks from the system (2.6).  
Setting  $v_x= Ay$, $v_y=Bx$ ($A$ and $B$ are constants to be
determined), and
${\bf v_A} = 0$, equations (2.6c,d) lead immediately to
$$
B(A- 2\Omega )= 2q\Omega^2, \qquad A(B+2\Omega) = 0
\eqno (2.8)
$$
There are two distinct branches of solutions.
One possibility is $A=0$, $B=-q\Omega$, which corresponds to simple
differential rotation of the background flow.  The other possibility is
$B= -2\Omega$, $A=\Omega(2-q)$.  This is epicyclic motion about the
guiding center $R=R_0$.  The streamlines are retrograde
ellipses with minor to major axis ratio $\sqrt{1-q/2}$.  We emphasize
the simple but important point that fluid elements in the neighborhood
of the corotation point do not separate from one another as time goes
on, but orbit instead about the local origin.  Therefore, in the
absence of Lorentz forces, magnetic field lines placed along the
epicyclic ellipses would remain undisturbed.

When $q=2$, the ellipses become degenerate, and shrink to straight
lines.  If $q>2$, this streamlines become hyperbolic, and correspond to
Rayleigh unstable flow (angular momentum decreases outward).

\bigskip
\centerline{2.2.2 \it{Elliptical Configurations.}}

When magnetic fields and Lorentz forces are self-consistently
included,
it is a remarkable fact that equilibrium solutions are still
possible.
Equations (2.6a--e) have the following exact solution:
$$
v_x = Ay, \quad v_y=Bx, \quad v_{Ax} = \alpha y, \quad v_{Ay} = \beta
x,
\eqno (2.9)
$$
where the constants $A, B, \alpha, \beta$ are given by
$$
B =\pm \Omega
\sqrt{ {2q\over 1 - \epsilon^2}}, \quad A/B=\alpha/\beta= - \epsilon^2
\eqno (2.10a)
$$
and 
$$
\alpha^2 = {2q\Omega^2\epsilon^4\over 1- \epsilon^4}\left(
1 \pm \sqrt{1-\epsilon^2\over q/2}\right).
\eqno (2.10b)
$$
Its physical significance of $\epsilon <1$ 
becomes evident from consideration of the equations for the flow
streamlines which coincide with the magnetic field lines:
$$
y^2 + {x^2\over\epsilon^2} = {\rm constant.}
\eqno (2.11)
$$
Thus, $\epsilon$ is the minor to major axis ratio of the elliptical
streamlines.  Unlike the hydrodynamical case, its value is not fixed by
the background $\Omega(R)$.  Furthermore, both retrograde {\it and\/}
prograde motion is possible; the former (latter) corresponds to taking
the $-$ ($+$) sign in equations (2.10a,b).  The sign of $\alpha$ has no
dynamical consequence (it determines the sense of the current flow),
but once chosen, $\beta$ must be of the opposite
sign.  Note that that the magnetic field lines do not shear, even
though a radial component is present and there is background (global)
differential rotation.  Rather, the field lines precisely track the
velocity streamlines, and rigidly rotate with the modified epicycle.
The ``miracle'' is that this behavior is dynamically fully
self-consistent.

Let us next consider the conditions under which retrograde or prograde
ellipses are obtained.  Retrograde motion corresponds to $A>0$, $B<0$,
i.e., to taking the minus sign in equations (2.10a,b).
The requirement that $\alpha^2>0$ means that
$$
1- q/2\le \epsilon^2 <1 \qquad {\rm (retrograde).}
\eqno (2.12)
$$ 
For a Keplerian disk, $\epsilon$ must therefore exceed $1/2$, for a
galactic $q=1$ law, $\epsilon$ must exceed $\sqrt{2}/2$.  The Keplerian
planet solutions of GNG, which were also retrograde ellipses, had a
complementary restriction on the domain $\epsilon$:  $0\le \epsilon\le 1/2$.
What is inaccessible hydrodynamically becomes available
magneto-hydrodynamically, and vice-versa.  Note that when
$1-\epsilon^2$ approaches $q/2$, we recover the the hydrodynamical
solution of the previous subsection.  This is what ultimately limits
the domain of $\epsilon$ for retrograde ellipses.  The inclusion of
magnetic hoop stresses ``plumps'' retrograde ellipses into more
circular structures.

Prograde motion corresponds to $A<0$, $B>0$ in equation (2.7).  The
prograde branch is not restricted by the condition $\alpha^2 >0$, since
this is obviously guaranteed by choosing the plus sign in equation
(2.10b).  Hence, the domain of $\epsilon$ for prograde ellipses, 
$$0\le \epsilon^2< 1 \qquad{\rm (prograde).}
\eqno(2.13)$$
extends beyond the retrograde domain.

The vorticity of the ellipses is given by
$$
B-A +2\Omega = \Omega\left[ 2 \pm (1+\epsilon^2)\sqrt{ {2q\over
1-\epsilon^2}}\right]
$$
Equation (2.12) implies that retrograde ellipses always have negative
vorticity; prograde ellipses have positive vorticity.

The Lorentz forces are $-\beta^2(1+\epsilon^2)x$ in the $x$ direction,
and $ - \epsilon^2\beta^2 (1+\epsilon^2)y$ in the $y$ direction.  This
is an inwardly directed force (confining) for both prograde and
retrograde ellipses, for a given $\epsilon$ 
larger in magnitude for the prograde solutions.
Prograde rotation has greater associated vorticity than retrograde
rotation, and larger confining forces are required.

\bigskip
\centerline{2.2.3 {\it Hyperbolic Configurations.}}

Hyperbolic configurations are another possible local solution in a disk.
Following the $AB\alpha\beta$ parameterization of equation (2.9), we obtain
$$
B=\pm \sqrt{    {2q\Omega^2\over1+ M^2 }  }, \quad A/B=\alpha/\beta =
M^2,
\eqno (2.14)
$$
and
$$
\alpha^2 = {2q\Omega^2M^4\over1-  M^4}\left(1\pm  \sqrt{1+M^2\over q/2}
\right)
$$
As before, the sign of $\alpha$ is not dynamically significant,
but once chosen,
$\alpha$ and $\beta$ must have the same sign.  
These solutions have flow streamlines and magnetic field configurations
given by the hyperbolae
$$
y^2-{x^2\over M^2} = {\rm constant.}
\eqno (2.15)
$$
Prograde hyperbolic flow corresponds to $B>0$, $M^2<1$; retrograde flow
to $B<0$, $M^2>1$.   (Note that with $\alpha^2$ is always positive
under these restrictions.)   Lorentz forces for hyperbolic flow are
directed neither radially inward nor outward (relative to our local
origin), but vary with quadrant.

\bigskip
\centerline{2.2.4 {\it Constant Density Solutions.}}

The planar field configurations of the previous section are fully
compatible with vertical stratification in the disk.  If one further
restricts the disk structure to the case of $\rho$ and $P$ being
independent of $z$, a greater variety of possible field configurations
exists.  This is not an uninteresting limit, because numerical
simulations are often performed ignoring the vertical disk structure,
and because the local midplane structure of astrophysical disks is
approximately one of constant density.  (More precisely, $\rho = \rho_0
- O(z^2)$, and we shall work to $O(z)$.)

We may first note that if $\rho$ and $P$ assumed to be constant,
nothing in our original solution would change if a constant vertical
magnetic field were present.  This is one possible generalization.  But
it is also possible to find qualitatively new solutions.

If $z$ components of the velocity and magnetic field are present, but
vertical structure is absent, instead of equation (2.6e),
our new $z$ equation is
$$
v_i\dd_i v_z = -\dd_z({v_A^2/2})+ v_{Ai}\dd_i v_{Az}
\eqno(2.16)
$$
Consider now a velocity of the form
$$ v_x= Az + D, \quad v_y = Bx, \quad v_z = Cx, \eqno
(2.17) $$
where $A,B,C$, and $D$ are constants, and an Alfv\'en velocity of the form 
$$ {\bf v_A} = \gamma
{\bf v} \eqno (2.18)
$$
with $\gamma$ constant.  The equations (2.6a,b) are automatically
satisfied, while equations (2.6c,d) and (2.16) lead to elliptical and
hyberbolic streamline solutions, as before.  Elliptical solutions
take the form
$$
C^2  = {2\Omega^2\over {1-\epsilon^2}}\left[2 (1+\epsilon^2)^2
-q\right], \quad C/A=-\epsilon^2, \quad \gamma^2={\epsilon^2\over 1+\eps^2},
\eqno(2.19a)
$$
and 
$$B=-2\Omega (1+\eps^2).
\eqno (2.19b)
$$
These solutions have the interesting property that $D$ is completely
unconstrained, and the entire ellipse can stream, in bulk, in the
radial direction!  Unusual streaming behavior is also characteristic of
the magnetorotational streaming solutions in two-dimensional
axisymmetric simulations (Hawley \& Balbus 1992).
However, these streaming solutions are
known to be unstable to Kelvin-Helmholtz types of instabilities when
three-dimensional structure is permitted (Goodman \& Xu 1994, Hawley,
Gammie, \& Balbus 1995).  The ordered flow breaks down into MHD
turbulence.

\bigskip
\sec{3. DISCUSSION}

The solutions presented in this paper represent both O-type
(elliptical) and X-type (hyperbolic) neutral points in the local
magnetic field topology.  In essence, we have shown that each of these
topologies has a range of well-defined field strengths that leads to
exact static equilibrium solutions to the local MHD equations in a
shearing disk.  It is remarkable that such solutions exist in the
presence of differential rotation.  

Unless $\epsilon^2 -1$ is very small, the local Alfv\'en velocities of
the solutions are of order $r\Omega$, where $r$ is the radial distance
from the local origin.  If $r \lta $ the disk scale height, the
magnetic field strengths are comparable to or less than thermal
values.  As $\epsilon \rightarrow 1$, field strengths becomes
unbounded, and the configurations are probably unstable.  At more
nominal field strengths, the question of stability is unclear.  Both
the Parker (1966) and magnetorotational (Balbus \& Hawley 1991)
instabilities are potentially disruptive, but in either case the
lengthscale and magnitude of the magnetic field make direct application
of the relevant criterion marginal.  The best option at this point is
direct numerical simulation.

But these concerns should not stifle speculation.  Even if these
structures prove unstable on rotational time scales, they may still
be astrophysically interesting.  X-type neutral points facilitate
reconnection, and their presence as an equilibrium configuration at
thermal field strengths may be an important saturation mechanism for
accretion disk dynamos.  Another application of interest is the
interface between an accretion disk and the magnetosphere of the
central source.  The magnetic field enforces rigid rotation with a
fiducial radius, the corotation radius, while differential rotation is
maintained outside corotation.  This work opens up the possibility
that magnetic fields at the interface need not be sheared!  The open
field lines of our hyperbolic solutions may globally join the
magnetospheric fields to the disk interior without producing large
off-diagonal Maxwell stresses.  

O-type elliptical structures offer interesting possibilities for
galactic disks.  If we identify these large scale vortices with sites
of molecular cloud complexes, there are a number of straightforward
predictions.   We should see, of course, at least a vaguely
elliptical morphology, with the major axis oriented azimuthally, and
magnetic fields following this morphology.  Both
retrograde and prograde rotation of the complexes is possible.  The
minor-to-major axis ratio of retrograde systems must exceed 0.7;
however prograde systems with considerably smaller values should exist.

Why have numerical simulations done to date not shown these
structures?   Three-dimensional MHD disk simulations have yet to reveal
the midplane elliptical and hyperbolic configurations.  There are
several possible reasons for this.  One, unlike the GNG solutions, the
density does not fall to zero at the edge of an ellipse (or at some
convenient point in an hyperbola).  This means these solutions must be
actively confined by the ambient disk, and conditions for this need not
arise spontaneously.  Two, numerical simulations have concentrated on
subthermal magnetic fields, because their presence leads to disk
turbulence.  If the field saturates at subthermal values, and it
generally does, the O and X neutral points may not be able to form.
(In the simulation where the traveling ellipses were seen, the field
strength grew to suprathermal values.) The best strategy might be to
start with thermal fields and allow some combination of differential
rotation, internal dynamical instabilities, and external driving to a
new field configuration.  Finally, there is always the possibility that
they structures never form spontaneously because they are too
unstable.

Undue pessimism, however, is not yet warranted.  Properly crafted
three-dimensional simulations should be able to clarify the most
important uncertainties.

\vskip 6pt

Part of this work was completed when one of us (SAB) was a visitor with
the radio astronomy group at the \'Ecole Normale Superieure, and he
would like to thank E.~Falgarone and M.~Perault for their generous
hospitality and advice.  This work has been supported by NASA grants
NAG5--4600, NAG5--3058, and by NSF grant AST--9423187.

\bigskip

\sec{REFERENCES}

\vskip 12 pt

\def\ref{\par\noindent\hangindent=0.5in\hangafter=1}

\ref Balbus, S.~A., \& Hawley, J.~F. 1991, ApJ, 376, 214
\ref Balbus, S.~A., \& Hawley, J.~F. 1998, Rev. Mod. Phys. 70, in press
\ref Goodman, J., Narayan, R., \& Goldreich, P. 1987, MNRAS, 225, 695
\ref Goodman, J., \& Xu, G. 1994, ApJ, 432, 213
\ref Hawley, J.~F. 1987, MNRAS, 225, 677
\ref Hawley, J.~F., \& Balbus, S.~A. 1992, ApJ, 400, 595
\ref Hawley, J.~F., Gammie, C.~F., \& Balbus, S.~A. 1995, ApJ, 440, 742
\ref Papaloizou, J.~C.~B., \&  Lin, D.~N.~C. 1995, ARAA, 33, 505
\ref Parker, E.~N. 1966, ApJ, 145, 811

\bye